\documentclass{article}

\usepackage{spconf}

\usepackage{amsbsy}			
\usepackage{amsmath}
\usepackage{amssymb}
\usepackage{amsfonts}
\usepackage{booktabs}
\usepackage{multirow}
\usepackage{latexsym}
\usepackage{amsthm}
\usepackage{ifthen}
\usepackage{tabularx}
\usepackage{csquotes}
\usepackage{lipsum}
\usepackage{bm}
\usepackage{algorithm}	
\usepackage{algorithmic}
\usepackage{graphicx,epstopdf}
\usepackage{caption}
\usepackage{subcaption}
\usepackage[normalem]{ulem}
\usepackage{xspace}
\usepackage{paralist}		
\usepackage{breqn}
\usepackage{cite}
\usepackage{color}
\usepackage[utf8]{inputenc}
\usepackage{textcomp}
\usepackage{arydshln}
\usepackage{graphicx,epstopdf}
\usepackage{caption}
\usepackage{subcaption}
\usepackage[normalem]{ulem}

\usepackage{xspace}
\usepackage{paralist}		
\usepackage{breqn}
\usepackage{cite}
\usepackage{color}

\allowdisplaybreaks

\newtheorem{remark}{Remark}

\let\oldremark \remark
\renewcommand{\remark}{\oldremark\normalfont}

\newcommand{\bx}{{\bm x}}

\newcommand{\bz}{{\bm z}}

\newcommand{\by}{{\bm y}}

\newcommand{\xf}{\widehat{\bx}_{\!_{i\!|\!i}}}
\newcommand{\xp}{\widehat{\bx}_{\!_{i\!+\!1\!|\!i}}}
\newcommand{\yf}{\widehat{\by}_{\!_{i\!|\!i}}}
\newcommand{\yp}{\widehat{\by}_{\!_{i\!+\!1\!|\!i}}}
\newcommand{\xpp}{\widehat{\bx}_{\!_{i\!|\!i\!-\!1\!}}}
\newcommand{\ypp}{\widehat{\by}_{\!_{i\!|\!i\!-\!1\!}}}

\title{Distributed Estimation of Dynamic Fields over Multi-agent Networks}
\name{Subhro~Das$^{\dagger}$,~Advisor:~Jos\'e~M. F.~Moura$^{\ddagger}$  
\thanks{\scriptsize This dissertation work~\cite{das2016thesis} has been submitted for PhD degree in Electrical and Computer Engineering at Carnegie Mellon University in June 2016. This work has been supported by NSF grants CCF1513936, CCF1011903 and CCF1018509. Simulations were run in Azure cloud supported by Microsoft Azure for Research Grant.}
}
\address{$^{\dagger}$ IBM T. J. Watson Research Center, Yorktown Heights, NY 10598, USA \\
	$^{\ddagger}$ ECE Department, Carnegie Mellon University, Pittsburgh, PA 15213, USA \\
               Email: subhro.das@ibm.com, moura@ece.cmu.edu}

\begin{document}

\ninept
\maketitle
\begin{abstract}
This work presents distributed algorithms for estimation of time-varying random fields over multi-agent/sensor networks. A network of sensors makes sparse and noisy local measurements of the dynamic field. Each sensor aims to obtain unbiased distributed estimates of the entire field with bounded mean-squared error (MSE) based on its own local observations and its neighbors' estimates. This work develops three novel distributed estimators: Pseudo-Innovations Kalman Filter (PIKF), Distributed Information Kalman Filter (DIKF) and {\it Consensus+Innovations} Kalman Filter (CIKF). We design the gain matrices such that the estimators achieve unbiased estimates with bounded MSE under minimal assumptions on the local observation and network communication models. This work establishes trade-offs between these three distributed estimators and demonstrates how they outperform existing solutions. We validate our results through extensive numerical evaluations.
\end{abstract}
\section{Introduction}
\label{sec:intro}
\vskip-5pt
{\bf Motivation:} In the era of large-scale systems and big data, distributed estimators, yielding robust and reliable state estimates by running local parallel inference algorithms, are capable of significantly reducing the large computation and communication load required by optimal centralized estimators, namely the Kalman-Bucy filter~\cite{kalman1961new}. Distributed estimators have applications in estimation, for example, of temperature, rainfall, or wind-speed over a large geographical area; dynamic states of a power grid; locations of a group of cooperating vehicles; or beliefs in social networks.  \\

\vskip-10pt
\noindent {\bf Related work:} Reference~\cite{mahmoud2013distributed} provides a bibliographic review of different approaches to distributed Kalman filtering and also discusses their applications. We classify the prior work on distributed dynamic field estimation into two groups, the second being further sub-divided in two sub-groups: i) Two timescale: Fast communication – slow dynamics and sensing; and ii) Single timescale: a)~Gossip Kalman filters; and b) Consensus+innovations estimators. The two timescale distributed estimators~\cite{olfati2009kalman} reproduce a Kalman filter locally at each sensor with a consensus step on the observations. The sensing and the local filter updates occur at the same slow time  scale of the process dynamics, while the consensus iterations happen at the fast time scale of communication among sensors. In single timescale estimators, sensing and communication occur once at every time step; the local estimators operate at the same time scale of the process dynamics. In gossip Kalman filters~\cite{li2015distributed}, the sensors exchange their field estimates and their error covariance matrices following the gossip protocol, whereas the single timescale estimator~\cite{khan2010connectivity} extends to time varying dynamics the {\it consensus$+$innovations} distributed estimator introduced in \cite{kar2012distributed} for parameter estimation. Here, we consider the single time scale setup because it reduces the onerous communication rounds required by the consensus step in the two timescale estimators. 
\section{Problem Formulation}
\label{sec:problem}
\vskip-5pt
\begin{figure}
	\centering
	\includegraphics[width=0.37\textwidth]{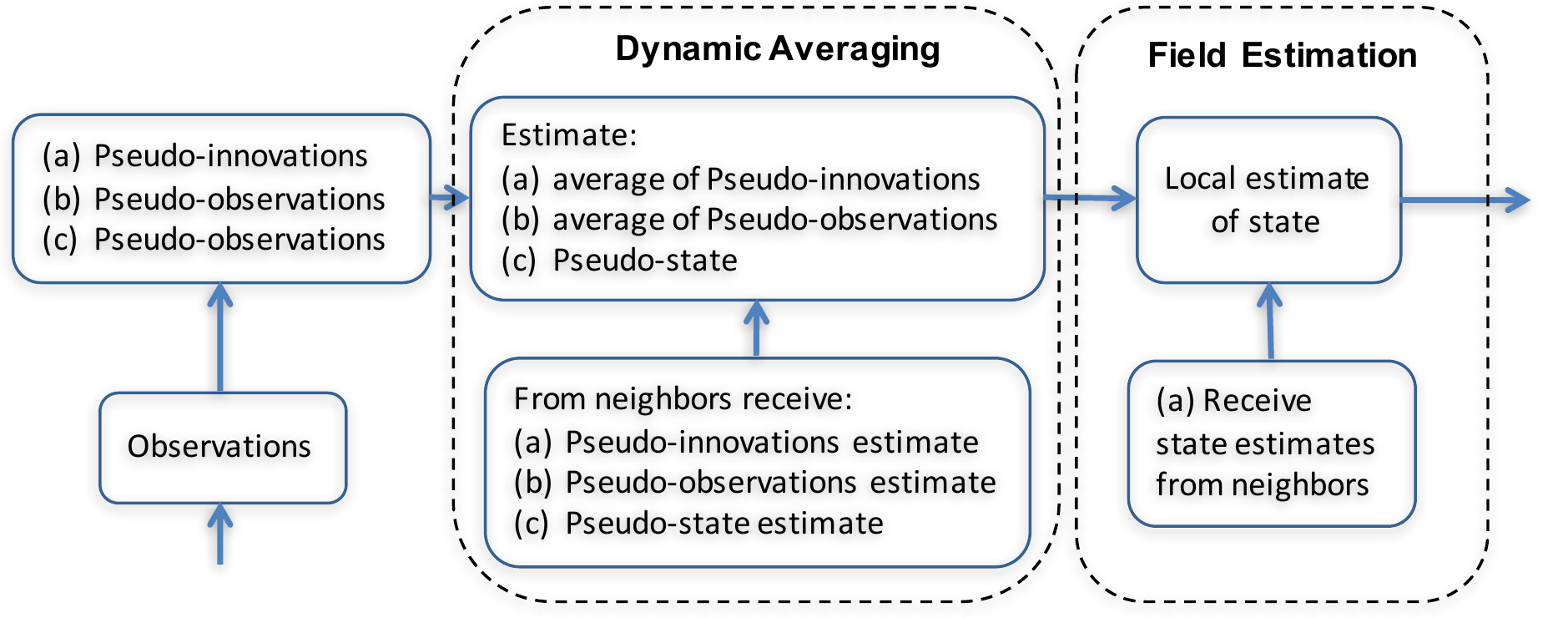}
	\caption{\scriptsize Structure of distributed estimators: (a) PIKF, (b) DIKF, and, (c) CIKF.}
	\label{fig:distributed_estimators}
	\hrule
	\vskip-17pt
\end{figure}
We consider a network of~$N$ sensors observing an underlying time-varying random field~$x_i~\in~\mathbb{R}^M$, where~$i$ is the time index. The evolution of the field~$x_i$ is:
\setlength\abovedisplayskip{4pt}
\setlength\belowdisplayskip{4pt}
\begin{align}
\label{eqn:x}
x_{i+1} = Ax_i + v_i
\end{align}
where, $A~\in~\mathbb{R}^{M\times M}$ is the field matrix and $v_i~\in~\mathbb{R}^{M}$ is the field noise. The noise~$v_i$ is zero-mean Gaussian, i.e., $v_i\sim\mathcal{N}(0,V)$. The initial condition, $x_0 \sim \mathcal{N}(\bar x_0,\Sigma_0)$, is normally distributed with mean $\bar x_0$ and covariance $\Sigma_0$. 

Let~$z^n_i \in \mathbb{R}^M$ denote the observation of the field~$x_i$ by the~$n^{\text{th}}$ sensor at time~$i$. The linear observations~$z^n_i$ follow
\begin{align}
\label{eqn:zn}
z^n_{i} = H_n x_i + r^n_{i}
\end{align}
where, $H_n \! \in \! \mathbb{R}^{M_n \times M}$ is a sparse observation matrix and~$r^n_{i} \! \in \! \mathbb{R}^{M_n}$ is the observation noise. In practice, the observation dimensions~$M_n,~\forall n$ are much smaller than the field dimension~$M$, i.e.,~$M_n \ll M$. The noise~$v_i$ is white Gaussian, i.e., $v_i\sim\mathcal{N}(0,V)$. The noise sequences~$v_i$,~$r^n_{i}$ and the field~$x_0$ are independent. The field is globally observable, but not necessarily locally observable. 

The sensor network is represented by an undirected, connected graph $\mathcal{G = (V,E)}$, where $\mathcal{V}$ and $\mathcal{E}$ denote the set of sensors and local communication channels respectively. The neighborhood of agent~$n$ is $\Omega_n = \{n\} \cup \{l|(n,l) \in  \mathcal{E}\}$. The Laplacian matrix of $\mathcal{G}$ is denoted by $L$, whose eigenvalues are $ 0 = \lambda_1(L) < \lambda_2(L) \leq . . . \leq \lambda_N(L)$. 
\section{Distributed estimators}
\label{sec:solution}
\vskip-5pt
In contradistinction with the centralized solutions, we identify that a key component in the distributed estimators that we develop is a global average step of a linearly transformed version of one of the three quantities: innovations, observations, or state. We refer to these linearly transformed and normalized versions of the local innovations, local observations, and state as pseudo-innovations, pseudo-observations, and pseudo-state, respectively. Apriori, it is not known which of these three pseudo-quantities will yield better performance, or what are the trade-offs in the computation complexity, or what are the communications constraints of each algorithm. Hence, we develop three distinct distributed estimators: $(i)$ Pseudo-Innovations Kalman Filter (PIKF in~\cite{das2013Allerton, das2013Asilomar, das2013EUSIPCO, das2013ICASSP}) using pseudo-innovations; $(ii)$ Distributed Information Kalman Filter (DIKF in~\cite{das2015TSP}) using pseudo-observations; and $(iii)$ {\it Consensus$+$Innovations} Kalman Filter (CIKF in~\cite{das2016TSP}) using pseudo-state. The structure of these estimators involves two steps -- a dynamic averaging step and a field estimation step, as shown in Fig.~\ref{fig:distributed_estimators}. The three versions of distributed estimators estimate the average of one of these local pseudo-quantities through a dynamic consensus step. The distributed field estimator uses this average estimate in the filtering step to compute the field estimates. Here, in Algorithm~1 we state only the CIKF~\cite{das2016TSP}, the readers are referred to \cite{das2015TSP, das2013Asilomar, das2013Allerton, das2013EUSIPCO, das2013ICASSP} for the PIKF and the DIKF. 

\begin{figure}
	\centering
	\includegraphics[width=0.37\textwidth]{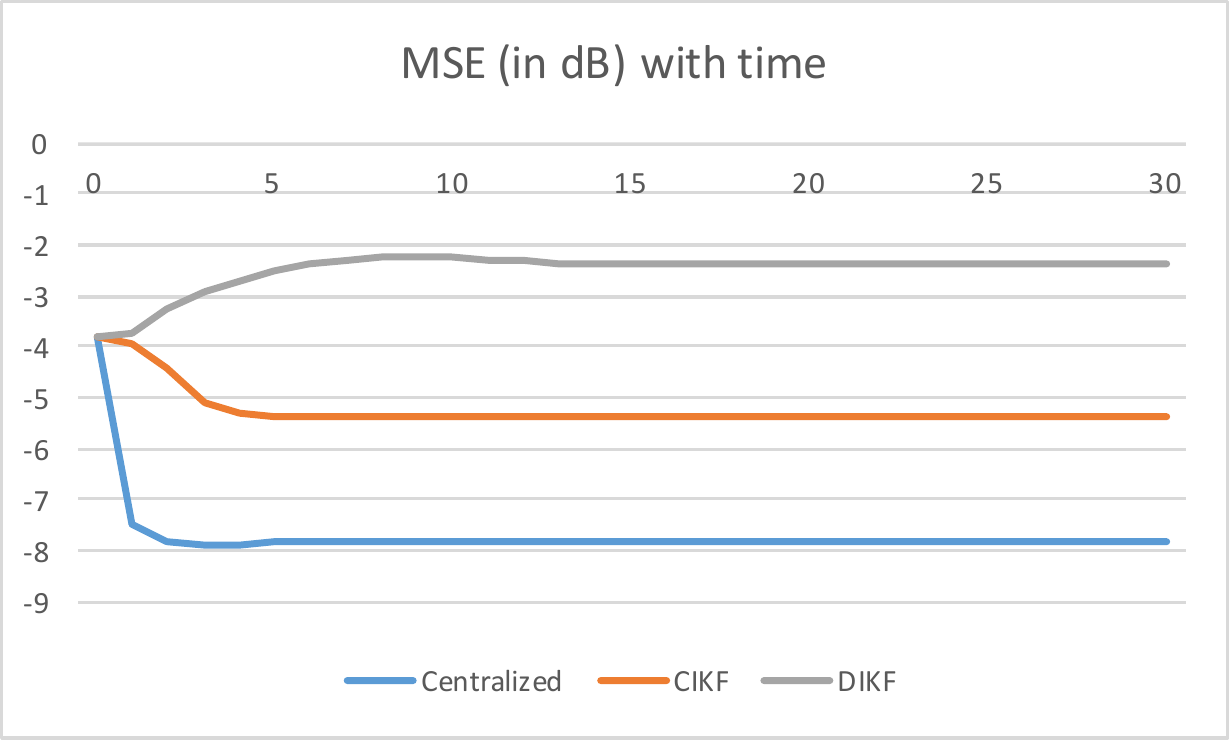}
	\caption{\footnotesize Comparison of MSE performance of the CIKF with CKF and DIKF.}
	\label{fig:mse_excel}
	\hrule
	\vskip-17pt
	\label{fig:sim}
\end{figure}

In distributed estimation of dynamic fields, while information diffuses through the network, the field itself evolves. This lag causes a gap in MSE performance between the distributed and centralized field estimators. Numerical simulations show that the CIKF performs better among the three in terms of MSE performance. The DIKF is 2dB better than the PIKF, see~\cite{das2015TSP}. As shown in Fig.~\ref{fig:mse_excel}, the CIKF improves the performance by 3dB over the DIKF, reducing by half the gap to the centralized (optimal) Kalman filter, while showing a faster convergence rate than the DIKF. 
\begin{algorithm}[t]
	\caption*{{\bf Algorithm:} \textit{Consensus$+$Innovations} Kalman Filter (CIKF)~\cite{das2016TSP}}
	\label{alg:algorithm}
	\begin{algorithmic}
		\STATE {\bfseries Input:} Model parameters $A$, $V$, $H$, $R$, $G$, $L$, $\overline{\bx}_0$, $\Sigma_0$.
		\vskip2pt
		\STATE {\bfseries Initialize:} $\widehat{\bx}^n_{0|-1} = \overline{\bx}_0$, $\widehat{\by}^n_{0|-1} = G\overline{\bx}_0$.
		\vskip2pt
		\STATE {\bfseries Pre-compute:} Gain matrices $B_i$ and $K_i$ using Algo.~2 in~\cite{das2016TSP}.
		\vskip2pt
		\WHILE{$i \geq 0$}
		\vskip2pt
		\STATE $\!\!\!\!${\textit{Communications:}}
		\vskip2pt
		\STATE Broadcast $\ypp^n$ to neighbors $\Omega_n$; receive $\{\ypp^l\}$ from $l \in \Omega_n$.
		\vskip2pt
		\STATE $\!\!\!\!${\textit{Observation:}}
		\vskip2pt
		\STATE Make measurement $\bz^n_i$; transform into pseudo-observation $\widetilde{\bz}^n_i$,
		\begin{footnotesize}
		\begin{align}
		\label{eqn:pseudo_obs}
		\widetilde{\bz}^n_i = H^T_n R_n^{-1} \bz^n_i, 
		\end{align}
	\end{footnotesize}
		\STATE $\!\!\!\!${\textit{Filter updates:} Compute $\yf^n$ and $\xf^n$}
		\begin{footnotesize}
		\begin{align}	
		\label{eqn:CIKF_yf}	
		\!\!\! \yf^n \!=\! \ypp^n &\! \!+\! \! \! \! \sum_{l \in \Omega_n}  \!\! \! B^{\!^{n\!l}}_i \! \! \left(  \!\ypp^l \!  \!\!-\! \ypp^n \right)   \!\!+\! \! B^{\!^{n\!n}}_i  \! \!\left(  \!\widetilde{\bz}^n_i  \!\!-\! \! \left( \!\widetilde{H}_n \ypp^n   \!\!\!+\! \check{H}_n \xpp^n  \!\right)  \!\right) \\
		\label{eqn:CIKF_xf}	
		\!\!\! \xf^n \!=\! \xpp^n & \!+\! K^{n}_i  \left( \yf^n \!-\! G \xpp^n \right) 
		\end{align}
		\end{footnotesize}
		\STATE $\!\!\!\!${\textit{Prediction updates:} Predict~$\yp^n$ and~$\xp^n$}
		\begin{footnotesize}
			\begin{align}	
			\label{eqn:CIKF_xyp}	
			\yp^n = \widetilde{A}\yf^n + \check{A}\xf^n, \qquad\qquad
			\xp^n = A \xf^n. \qquad\qquad\qquad \;\;
			\end{align}
		\end{footnotesize}
		\ENDWHILE
		\vskip-50pt
	\end{algorithmic}
\end{algorithm}

\vskip-50pt

\section{Contributions}
\label{sec:conclusion}
\vskip-5pt
The contributions of this work are: ${\bf(a)}$~Proposed novel {\it Consensus $+$ Innovations} distributed estimators~\cite{das2016TSP, das2015TSP, das2013Asilomar, das2013Allerton, das2013EUSIPCO, das2013ICASSP}, PIKF, DIKF, and CIKF of dynamic fields over multi-agent networks and proved their convergence under minimal assumptions on the local observation and network communication models;  ${\bf(b)}$ Introduced the relevance of pseudo-innovations~\cite{das2013Allerton, das2013Asilomar, das2013EUSIPCO, das2013ICASSP}, pseudo-observations~\cite{das2015TSP}, and pseudo-state~\cite{das2016TSP} in the context of distributed field estimation, and, developed a  {\it Consensus$+$Innovations} distributed dynamic averaging algorithm as a key part of field estimation; ${\bf(c)}$~Designed the optimized filter gain matrices~\cite{das2016TSP} using the Gauss-Markov theorem, so that the estimation MSE is minimized, and, derived the distributed version of the algebraic Riccati equation for the DIKF and the CIKF~\cite{das2016TSP, das2015TSP}; ${\bf(d)}$~Expressed the tracking capacity of the distributed estimators~\cite{das2016TSP, das2015TSP, das2013Asilomar, das2013Allerton, das2013EUSIPCO, das2013ICASSP} in terms of the network Laplacian and the observation matrices; ${\bf(e)}$~Validated the convergence results~\cite{das2016TSP, das2015TSP, das2013Asilomar, das2013Allerton, das2013EUSIPCO, das2013ICASSP} through numerical simulations and evaluated experimentally the sensitivity of the performance of the DIKF~\cite{das2015TSP} with respect to model parameters, noise statistics, and network models. \\

\vskip-5pt
\noindent {\bf Future work:} The generic nature of the results in this thesis makes them applicable in a variety of problems in distributed inference. Interesting extensions include developing a distributed estimator that is resilient to random sensor link and node failures. This is important in practice, because the communication links between sensors or the sensors themselves can fail. For distributed parameter estimation, this has been considered in~\cite{kar2012distributed}.

\begin{footnotesize}
\bibliographystyle{IEEEbib}
\bibliography{reference}
\end{footnotesize}

\end{document}